\documentclass[a4paper]{revtex4}
\usepackage{graphicx}
\usepackage{fancyhdr}
\usepackage{amsmath,multirow}
\usepackage{color}

\begin{document}

\title{New method for precise determination of top quark mass at LHC}

%

\author{Sayaka Kawabata}
\affiliation{Department of Physics, Tohoku University, Sendai 980-8578, JAPAN}

\begin{abstract}
Current measurements of the top quark mass which have achieved a precision of less than $1$\,GeV involve a theoretical problem that the definition of the measured mass is ambiguous in perturbation theory.
As a possible solution to the problem, we present a new method to measure the top quark mass at the LHC.
This method uses lepton energy distribution and has a boost-invariant nature.
We discuss strategies towards a precise determination of theoretically well-defined top quark masses such as the $\overline{\rm MS}$ mass with the method.
As a first step in this direction, a simulation analysis at the leading order is performed considering actual experimental circumstances.
The result indicates that this method with further improvements is capable of realizing a precision of less than $1$\,GeV at the LHC.
\end{abstract}

\maketitle

\thispagestyle{fancy}


\section{Introduction}
\label{sec:introduction}
The top quark mass is a key parameter in various particle physics.
It is an important input to electroweak precision fits, and 
its precise value is required for the consistency check of the Standard Model (SM) and constraining models of new physics with these fits~\cite{Baak:2012kk,Ciuchini:2013pca,Baak:2014ora}.
In addition, the top quark mass plays a decisive role in examination of the SM vacuum stability~\cite{Degrassi:2012ry,Buttazzo:2013uya}.
The top quark mass has been measured at the Tevatron and at Run I of the Large Hadron Collider (LHC).
Their first combined result yields $m_t=173.34\pm 0.76$\,GeV~\cite{ATLAS:2014wva}.
Higher precision is expected to be achieved by the upgraded LHC with the collision energy of $\sqrt{s}=13$-$14$\,TeV~\cite{CMS:2013wfa}.

However, it has been recognized that the definition of the measured mass is ambiguous in perturbation theory.
The measured mass is often referred to as ``Monte-Carlo (MC) mass" and distinguished from theoretically well-defined top quark masses, e.g. the $\overline{\rm MS}$ mass and the pole mass.
The reason for this is as follows:
in the above measurements, the top quark mass is basically obtained from measuring kinematic distributions of the top quark final state and comparing the distributions with theoretical predictions.
For the theoretical predictions, MC event generators are used.
Although the predictions ought to be based on reliable perturbation theory to extract theoretically well-defined top quark masses,
formation of jets which includes hadronization processes cannot be derived from first principles within the framework of perturbative QCD.
For this reason, phenomenological models are used to describe hadronization in the MC generators.
As a result, the top quark mass obtained using jet momenta depends on hadronization models which are not based on perturbation theory.
Note that we cannot separate decay products purely from a top quark in the first place.
The top quark inevitably has color connections to other colored particles, and thus, hadronization always involves other particles, too.
Moreover, process dependence of hadronization makes it difficult to quantify uncertainties of the models further.
More detailed discussion about this problem can be found in, e.g. Ref.~\cite{Hoang:2014oea}.

One way to avoid this problem is to use inclusive observables instead of kinematics of top quarks.
The ATLAS and the CMS collaborations have obtained the top quark pole mass as $m_t^{\rm pole}=172.9^{+2.5}_{-2.6}$\,GeV~\cite{Aad:2014kva} and $m_t^{\rm pole}=176.7^{+3.8}_{-3.4}$\,GeV~\cite{Chatrchyan:2013haa}, respectively, from measurements of inclusive $t\overline{t}$ production cross section.
With the same approach, the $\overline{\rm MS}$ mass has also been obtained as $m_t^{\overline{\rm MS}}(m_t^{\overline{\rm MS}})=160.0^{+5.1}_{-4.5}$\,GeV by the D0 collaboration~\cite{Abazov:2011pta}.
Although these masses have clear definitions in perturbation theory, their errors are still considerably large.

If one aims at a high accuracy of less than $1$\,GeV in determination of theoretically well-defined top masses, we must take care of the fact that the quark pole mass suffers from an intrinsic uncertainty of the order of $\Lambda_{\rm QCD}$.
A quark has a color, and therefore, we cannot extract a single quark.
It means that in principle the top quark propagator does not have a pole if it is computed non-perturbatively.
Therefore, we can specify the top quark pole mass only by defining it order-by-order in perturbation theory.
Though the pole mass is useful in some cases, it is sensitive to infrared physics and exhibits poor convergence of the perturbative series when used in perturbative computations.
Avoiding this difficulty requires utilization of the so-called short-distance masses, which improves convergence of perturbation drastically.
Among others, the $\overline{\rm MS}$ mass is commonly used, where improved convergence has been confirmed in various perturbative QCD predictions~\cite{Langenfeld:2009wd,Hoang:1998nz,Dowling:2013baa}. 
In this context, the top quark $\overline{\rm MS}$ mass, among various definitions of the top quark mass, should be determined with high precision.

In the light of the above situation and theoretical issues, we propose a new method for top quark mass reconstruction aiming a precise determination of the top quark $\overline{\rm MS}$ mass at the LHC~\cite{Kawabataa:2014osa}.
The method utilizes lepton energy distribution.
Since the lepton in the top decay $t\rightarrow  bW\rightarrow b\ell\nu$ is not affected by hadronization effects, our method can determine theoretically well-defined top quark masses.
Moreover, another important feature of our method is a boost invariance.
The lepton energy distribution depends on velocity distribution of parent top quarks, which is difficult to know accurately due to, e.g. uncertainties in parton distribution functions (PDFs).
On the other hand, our method is, in principle, independent of top quark velocities.
The experimental observable we use in this method is lepton energy distribution in any Lorentz frame, and theoretical prediction compared with it is the lepton energy distribution in the rest frame of the top quark.
In this comparison, the boost dependence of the distribution is canceled, which is a non-trivial point of our method.
We call this method the ``weight function method."

In Sec.~\ref{sec:mtDetermination} we present an outline of the weight function method and strategies towards a precise determination of the top quark pole and $\overline{\rm MS}$ masses with the method.
As a first step in the direction, we perform a simulation analysis at the leading order (LO), investigating the experimental viability of the method.
The result of the LO analysis and its consequent prospects are given in Sec.~\ref{sec:LOAnalysis}.

\section{Top mass determination with weight function method}
\label{sec:mtDetermination}
In this short article, we give only the main points of the weight function method.
For further details of the method we refer to Refs.~\cite{Kawabata:2011gz,Kawabata:2013fta,Kawabataa:2014osa}.
In the weight function method, the top quark mass can be obtained by the following three steps:
\begin{enumerate}
	\item Compute a weight function $W(E_\ell,m)$ which is given by
		\begin{equation}
			W(E_\ell,m)=\!\int \!dE \left.\mathcal{D}_0(E;m)\frac{1}{EE_\ell} \,({\rm odd~fn.~of~}\rho)\right|_{e^{\rho}=E_\ell/E}\,,
			\label{eq:WF}
		\end{equation}
		where $\mathcal{D}_0(E;m)$ is the normalized lepton energy distribution in the rest frame of the top quark whose mass is $m$.
	\item Construct a weighted integral $I(m)$, using the weight function $W(E_\ell,m)$ and a measured lepton energy distribution $D(E_\ell)$:
		\begin{equation}
			I(m) ~\equiv~ \int dE_\ell \,D(E_\ell) \,W(E_\ell,m)\,.
		\end{equation}
	\item Then the top quark mass can be obtained as the zero of $I(m)$:
		\begin{equation}
			I(m=m_t^{\rm rec}) ~=~ 0\,.
		\end{equation}
\end{enumerate}
The method is based on an assumption that the lepton angular distribution in the top quark rest frame is flat~\footnote{
For top quarks produced via $t\overline{t}$ at the LHC, this assumption is valid at sub-percent level~\cite{Bernreuther:2006vg}.
The deviation from the flatness should be taken into account as small corrections if needed.
}.
Note that this method does not consider finite-width effects of the top quark.
The effects should be incorporated as small corrections to the method.

For realization of a precise top mass determination with the method, experimental viability of the method is vital.
We have confirmed this in Ref.~\cite{Kawabataa:2014osa}, by performing a simulation analysis at LO with consideration of actual experimental circumstances.
The results are summarized in the following section.
The subjects of our future works will be to include theoretical corrections such as higher-order QCD corrections and finite-width effects of the top quark.
In the rest of this section, we discuss strategies towards a precise determination of the top quark pole and $\overline{\rm MS}$ masses considering these theoretical corrections.


With the weight function method, the top quark pole mass can be obtained by computing the lepton energy distribution in the top quark rest frame $\mathcal{D}_0(E;m)$ in the on-shell scheme.
In the computation of $\mathcal{D}_0$, the pole mass parameter of the top quark is supposed to be set to $m$.
Then the zero of $I(m)$ gives the value of the top quark pole mass.
Higher-order corrections which are relevant to $\mathcal{D}_0$ are only those concerning the top-quark decay process (ignoring effects of the top quark width
), and
they are now available to next-to-next-to-leading order (NNLO) in perturbative QCD~\cite{Gao:2012ja,Brucherseifer:2013iv}.
The obtained pole mass can be converted to the $\overline{\rm MS}$ mass using the relation between them which is currently complete to four-loop order~\cite{Marquard:2015qpa}.

As we have mentioned in the previous section, the quark pole mass suffers from intrinsic ambiguities and leads to a bad convergence behaviour of the perturbative series.
Consequently, the determination of the $\overline{\rm MS}$ mass via the pole mass would receive uncertainties.
A naive approach to obtain the $\overline{\rm MS}$ mass directly (namely bypassing the pole mass) with the weight function method is to compute $\mathcal{D}_0$ in the $\overline{\rm MS}$ scheme, setting the $\overline{\rm MS}$ mass parameter of the top quark to $m$.
Though our method is not limited to a particular renormalization scheme, the lepton distribution in the top quark rest frame $\mathcal{D}_0$ is not suited to the $\overline{\rm MS}$ scheme in a certain region of phase space if we naively perform $\alpha_s$ expansion~\footnote{See similar discussions in Refs.~\cite{Beneke:1998rk,Falgari:2013gwa}.}.
Therefore, it would be helpful to use other short-distance masses which differ from the pole mass by smaller amounts than the $\overline{\rm MS}$ mass as well as being insensitive to infrared physics.
Such masses are, for example, the potential-subtracted (PS) mass~\cite{Beneke:1998rk} and the 1S mass~\cite{Hoang:1999zc}.
Once we obtain these masses with our method, they can be converted to the $\overline{\rm MS}$ mass.
In addition, since the weighted integral $I(m)$ is conceptually close to an inclusive observable, being integrated over the lepton energy, the possibility that $I(m)$ [instead of $\mathcal{D}_0(E;m)$] is a suitable observable for naive $\alpha_s$ expansion in the $\overline{\rm MS}$ scheme is worth investigating.
 
Furthermore, consideration of finite-width effects of the top quark is necessary to achieve a high precision.
The weight function method accommodate the so-called factorizable corrections
in resonant and sub-resonant contributions if there exists an intermediate top quark which emits a lepton.
On the other hand, the method does not consider effects due to off-shellness (propagation corrections), non-factorizable corrections and sub- or non-resonant contributions where no parent top quark of a lepton intermediates.
Corrections required due to these effects can be computed based on such frameworks as the complex-mass scheme~\cite{Denner:1999gp} and the effective field theory (EFT) approach~\cite{Beneke:2003xh,Falgari:2013gwa}.

\section{Simulation analysis at LO}
\label{sec:LOAnalysis}

In this section, we outline the results of a simulation analysis of the top mass reconstruction with the weight function method at LO, which is performed in Ref.~\cite{Kawabataa:2014osa}.
In the analysis, we investigate experimental viability of the method, considering various experimental effects.
We study $t\overline{t}$ production events with lepton+jets final states at $\sqrt{s}=14$\,TeV at the LHC.

Among various sources of experimental effects, such as detector acceptance, event selection cuts and background contributions, lepton cuts affect our method most seriously.
Typical lepton cuts at the LHC are, for example, $p_T(\ell)>20$\,GeV and $|\eta (\ell)|<2.4$.
These cuts severely deform lepton distributions, and as a consequence, the obtained mass with our method (the zero of $I(m)$) deviates from the true mass value.
This requires a modification of the method and we cope with it by compensating the lost events by the lepton cuts with MC simulation events.
The true mass can be obtained as the zero of $I(m)$ imposing the consistency condition that the top mass value $m_t^c$ of the compensated MC part coincides with the zero of $I(m)$.
The normalization of lepton distributions of the compensated part is determined so that its junction to the data part is smooth.
We confirmed that the compensated MC part is less sensitive to the top mass determination owing to this way of normalization determination.
Uncertainties of the compensated MC part associated with factorization scale dependence and PDF uncertainties are estimated.

\begin{figure}[t]
	\begin{center}
		\begin{tabular}{c}
		
		 	\begin{minipage}{0.45\hsize}
					\includegraphics[width=.8\textwidth]{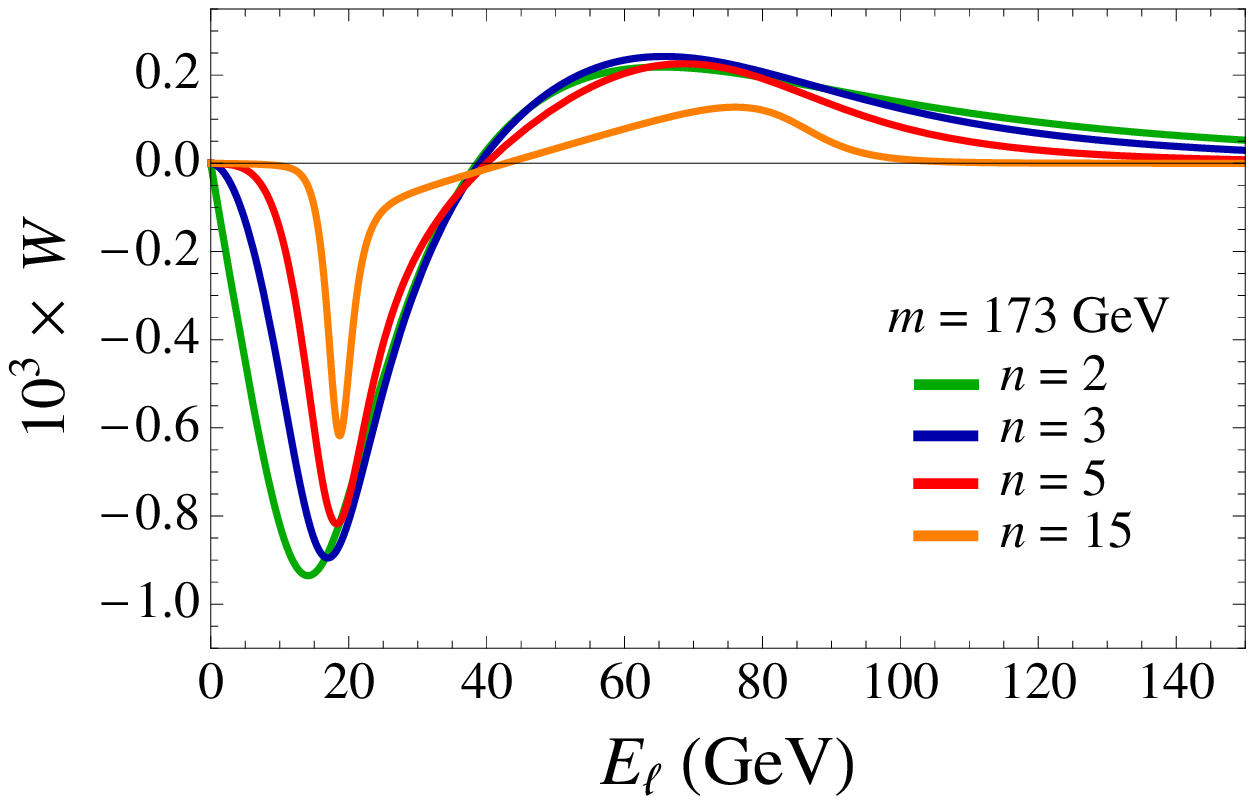}
					\caption{\label{fig:WFn=2-15} Weight functions $W(E_\ell,m)$ at LO with the odd function of $\rho$ in Eq.~(\ref{eq:WF}) taken to be $n\,\tanh (n\rho)/\cosh (n\rho)$.}
			\end{minipage}
\hspace{1cm}	
		 	\begin{minipage}{0.45\hsize}
					\includegraphics[width=.8\textwidth]{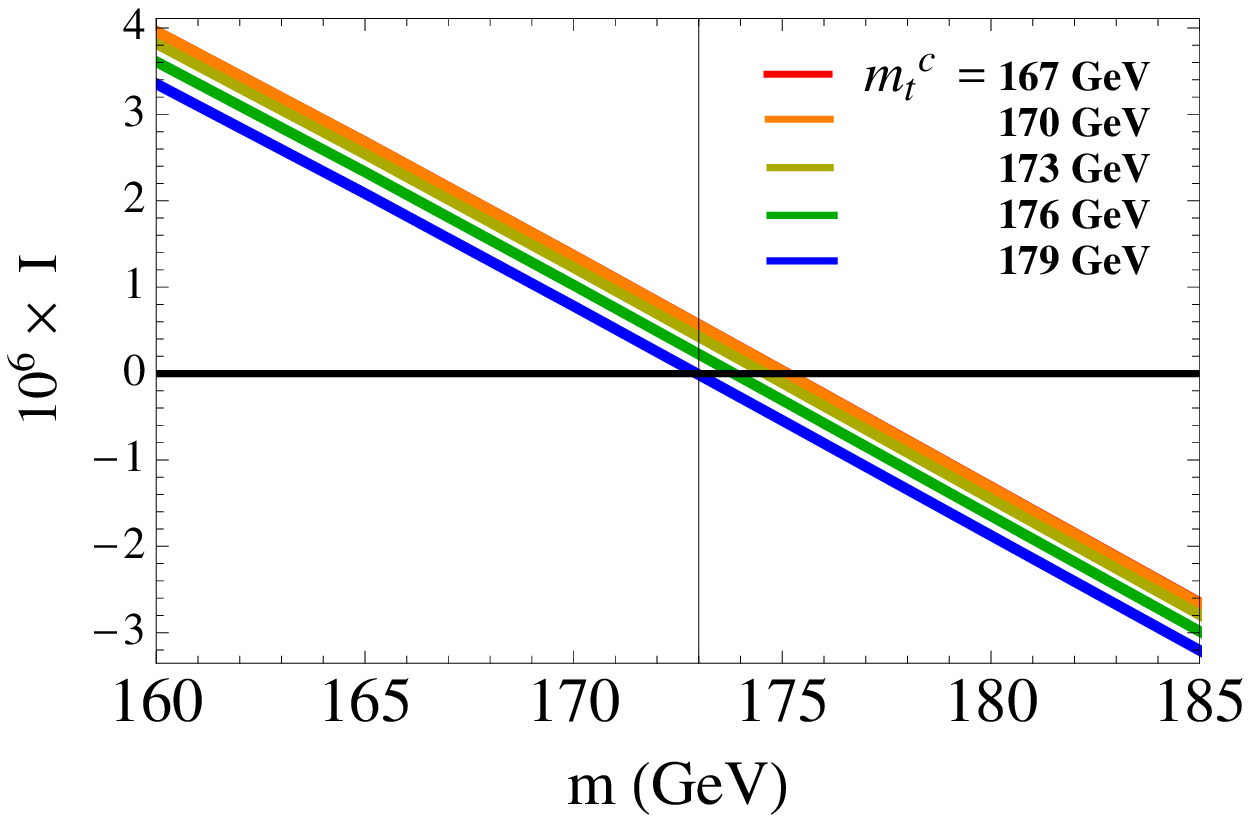}
					\caption{\label{fig:ImWithmtMCAftCuts} Weighted integrals $I(m)$ with various $m_t^{\rm c}$. The weight function used corresponds to $n=2$. The input value of the top quark mass is $173$\,GeV.}
			\end{minipage}		
						
		\end{tabular}	
	\end{center}
	\vspace{-.5cm}
\end{figure}

In Fig.~\ref{fig:WFn=2-15} we show weight functions $W(E_\ell,m)$ at LO with the odd function of $\rho$ in Eq.~(\ref{eq:WF}) taken as $n\,\tanh (n\rho)/\cosh (n\rho)$.
Using the weight functions and the compensated lepton distributions with various $m_t^c$ after event selection cuts, we construct weighted integrals $I(m)$.
Fig.~\ref{fig:ImWithmtMCAftCuts} shows the weighted integrals with the weight function corresponding to $n=2$.
One can see in Fig.~\ref{fig:ImWithmtMCAftCuts} that the variation of the zero of $I(m)$ is much less than the corresponding variation of $m_t^c$.
The input value of the top quark mass to the data part, which corresponds to the true mass in real experiments, is $173$\,GeV in the figure.
From these weighted integrals, we reconstruct the top quark mass as $174.1$\,GeV in this simulation analysis.
The estimated MC statistical error due to limitation of the capable number for generated and analyzed events is $+1.0$/$-1.1$\,GeV.
In addition, the expected shift due to effects of the top width is $+0.34$\,GeV.
Taking them into account, the reconstructed mass is consistent with the input mass.
Table~\ref{tab:uncertainties} summarizes estimates of uncertainties from several major sources in the top mass determination at LO.
\begin{table}[t]
	\centering
	\caption{Estimates of uncertainties in GeV from several sources in the top mass determination at LO. The weight function used in this evaluation corresponds to $n=2$. The signal and background statistical errors correspond to those with an integrated luminosity of $100$\,fb$^{-1}$.}
	\vspace{.06cm}
	\begin{tabular}{c|c|c|c|c}
		~Signal stat. error~ & ~~~~~Fac. scale~~~~~ & ~~~~~~~PDF~~~~~~~ & ~Jet energy scale~ & ~Background stat. error~ \\ \hline
		$0.4$ & $+1.5$/$-1.4$ & $0.6$ & $+0.2$/$-0.0$ & $0.4$\\
	\end{tabular}
	\label{tab:uncertainties} 
	\vspace{-0.1cm}
\end{table}

In ideal experiments, the weight function method does not depend on uncertainties concerning top-quark production processes, such as factorization scale dependence and PDF uncertainties (as stated in Sec.~\ref{sec:introduction}).
In practice, however, it depends on these uncertainties via the compensated part, and as a result, the factorization scale uncertainties are the dominant source in the LO analysis.
These uncertainties are expected to be reduced by using a MC generator with NLO corrections to top production processes for the compensated part.
The result of the LO analysis indicates that with further improvements mentioned in Sec.~\ref{sec:mtDetermination}, this method is capable of realizing a precision of less than $1$\,GeV in the determination of the top quark $\overline{\rm MS}$ mass at the LHC.


\begin{acknowledgments}
This work is based on analyses performed in collaboration with Y. Shimizu, Y. Sumino and H. Yokoya, and I would like to thank all of them.
The work was supported by Grant-in-Aid for JSPS Fellows under the program number 24$\cdot$3439.
\end{acknowledgments}

\bigskip 

\end{document}